\documentstyle[emulateapj,psfig]{article}
\newcommand{\be}{\begin{equation}}
\newcommand{\ba}{\begin{eqnarray}}
\newcommand{\ee}{\end{equation}}
\newcommand{\ea}{\end{eqnarray}}  
\newcommand{\etal}{et al.\ }
\def\gtsima{$\; \buildrel > \over \sim \;$}
\def\ltsima{$\; \buildrel < \over \sim \;$}
\def\gsim{\lower.5ex\hbox{\gtsima}}
\def\lsim{\lower.5ex\hbox{\ltsima}}
\def\simgt{\lower.5ex\hbox{\gtsima}}
\def\simlt{\lower.5ex\hbox{\ltsima}}
\def\simpr{\lower.5ex\hbox{\prosima}}

\def\msun{{M_\odot}}

\def\E3{{\cal E}_{\rm g}^{III}}

\begin{document}
\title{On the Spatial Correlations of Lyman Break Galaxies}

\author{Evan Scannapieco\altaffilmark{1} \& Robert J. 
Thacker\altaffilmark{2}}
\altaffiltext{1}{Osservatorio Astrofisico di Arcetri, Largo E.
  Fermi 5 Firenze, Italy}

\altaffiltext{2}{Department of Physics and Astronomy,
McMaster University, 1280 Main St.\ West, Hamilton, Ontario, L8S 4M1,
Canada.}

\begin{abstract}

Motivated by the observed discrepancy between the strong spatial
correlations of Lyman break galaxies (LBGs) and their velocity
dispersions, we consider a theoretical model in which these
starbursting galaxies are associated with dark matter halos that
experience appreciable infall of material.  We show using numerical
simulation that selecting halos that substantially increase in mass
within a fixed time interval introduces a ``temporal bias'' which
boosts their clustering above that of the underlying population.  If
time intervals consistent with the observed LBGs star formation rates
of $\sim 50 \msun$ yr$^{-1}$ are chosen, then spatial correlations are
enhanced by up to a factor of two.  These values roughly correspond to
the geometrical bias of objects three times as massive.  Thus, it is
clear that temporal biasing must be taken into account when
interpreting the properties of Lyman break galaxies.

\end{abstract}


\section{Introduction}

Cosmologists love to count and correlate, and mostly for good reasons.
If structure formation proceeds through gravitational instability,
then the number densities and distributions of cosmological objects
are directly dependent on the underlying physical model.  Thus
measuring the number of objects as function of virial velocity and
redshift provides a direct probe of the primordial power spectrum and
the overall cosmological parameters.  This technique has been applied
most cleanly to galaxy clusters, whose number densities and evolution
provide strong constraints on the overall matter density (eg., Eke,
Cole, \& Frenk 1996; Bahcall, Fan, \& Cen 1997; Carlberg, Yee, \&
Ellingson 1997).

Similarly, because the density peaks in which objects form are more
clustered than the underlying mass distribution, the clustering 
of virialized structures provide a wealth of information.  This
``geometrical bias'' is a systematic function of the mass of these
structures, an effect that has been well-studied analytically and
numerically (Kaiser 1984; Bardeen et al. 1986; Mo \& White 1996;   
Porciani \etal 1998; Jing 1999; Scannapieco \& Barkana 2002).  

Perhaps the most famous measurement of this biasing has been in the
large sample of $z \sim 3$ galaxies made available by the Lyman-break
color-selection technique (Adelberger \etal 1998; Steidel \etal 1998).
Early papers pointed out that if the mass of these Lyman break
galaxies (LBGs) could be determined, their clustering would serve as a
sensitive test of cosmology (Mo \& Fukugita 1996; Adelberger et al.\
1998; Giavalisco et al.\ 1998). Later efforts inverted this approach,
using a wide range of cosmological constraints to compute the bias of
LBGs and relate this to their overall mass (eg.\ Coles et al.\ 1998;
Giavalisco \& Dickenson 2001; Shu, Mao, \& Mo 2001; Wechsler et al.\
2001; Porciani \& Giavalisco 2002).  Such comparisons have shown that
the clustering of LBGs brighter than ${\cal R}_{AB} \leq 25.5$ is
roughly that expected from the geometrical bias of $10^{12} \msun$
objects in the currently favored cosmological model. Furthermore, the
theoretical number density of $10^{12} \msun$ objects at $z = 3$ is
consistent with the observed densities of LBGs, roughly $2 \times
10^{-3}h^{3}$ Mpc$^{-3}$, where $h$ is the Hubble constant in units of
100 km s$^{-1}$ Mpc$^{-1}$.  Thus counting and correlating suggest a
one-to-one correspondence between LBGs and $10^{12} \msun$ objects.

Yet there are problems with this simple picture.  LBGs are extremely
luminous in their rest-frame UV, implying star formation rates on the
order of $\sim 50 \msun$ per year (eg.\ Adelberger \& Steidel 2000).
Thus a one-to-one correspondence means that {\em all} $10^{12} \msun$
objects must be forming stars at an enormous rate at $z = 3$.
Furthermore, the linewidths measured from the nebular emission of a
spectroscopic sample of the brightest of such galaxies gives projected
velocity dispersions of $50-115$ km/s (Pettini et al.\ 2001), which
correspond to total masses $\leq 10^{11} M_\odot$, if interpreted as
circular velocities.

In this {\em Letter} we explore an alternative possibility.  We
associate LBGs with a limited subset of objects that experience an
appreciable increase in mass, which we naturally associate with a
starburst.  Through a detailed numerical simulation we show that such
accreting groups are more clustered than the general population,
mimicking the properties of higher-mass halos, and modifying the mass
scales that are most naturally associated with LBGs.

This ``temporal biasing'' has never before been measured in
simulations, and is not dependent on any merger criteria or properties
of the accreted material.  However the idea that LBGs correspond to
merger induced starbursts has been proposed (eg.\ Kolatt \etal 1999),
and several authors have conducted numerical studies of the bias of
mergers, obtaining mixed results.  Kauffmann \& Haehnelt (2002)
analyzed the cross-correlation between objects undergoing
major-mergers and the general population, finding weak enhancement at
small distances.  Gott\"ober et al.\ (2002) found that applying a
merger criterion at $z=1$ can effect the bias of objects at $z = 0.$
Finally, Percival et al.\ (2003, hereafter P03) applied a set of
merger criteria at $10^8$ yr intervals, failing to obtain enhancement
as discussed in detail below.

The structure of this work is a follows:  In \S 2 we describe our
numerical simulation and in \S 3 we discuss our group-finding
algorithms and develop a robust definition of accreting groups.
In \S 4 we present our results for the correlation functions of these
samples, and a discussion is given in \S 5.

\section{Simulations}

Driven by measurements of the Cosmic Microwave Background, the number
abundance of galaxy clusters, and high redshift supernova distance
estimates (eg.\ Spergel et al.\ 2003; Eke \etal 1996;
Perlmutter \etal 1999) we focus our attention on a Cold Dark Matter
cosmological model with parameters $h=0.7$, $\Omega_0$ = 0.3,
$\Omega_\Lambda$ = 0.65, $\Omega_b = 0.05$, $\sigma_8 = 0.87$, and
$n=1$, where $\Omega_0$, $\Omega_\Lambda$, and $\Omega_b$ are the
total matter, vacuum, and baryonic densities in units of the critical
density, $\sigma_8^2$ is the variance of linear fluctuations on the $8
h^{-1}{\rm Mpc}$ scale, and $n$ is the ``tilt'' of the primordial
power spectrum. The Bardeen et al.\ (1986) transfer function was used
with an effective shape parameter of $\Gamma=0.18$.

The two competing desires of achieving high mass resolution while
simulating a large sample of halos led us to use a box size of $73$ 
comoving Mpc on a side, populated with $350^3$ dark matter
particles.  The mass of each particle was $4.3 \times 10^{8}
\msun$, which gives a nominal minimum mass resolution for our group
finding of $3.4 \times 10^{10} \msun$ as we select only groups
with 80 or more particles. The simulation was started at an initial
redshift of $z=49$, and a fixed physical Plummer softening length of
5.7 kpc was chosen. The simulations used a
parallel OpenMP-based version of the HYDRA code (Couchman \etal 1995,
Thacker \& Couchman 2000), with 64-bit precision being used
throughout.

\section{Group Finding and Halo Tracing}

Group finding is a widely studied topic in cosmology since it can lead
to (small) systematic differences (eg Jenkins \etal 2001). To
demonstrate the robustness of our results we have chosen two distinct
group finding approaches; the friends-of-friends approach (Davis \etal
1985, FOF) and the HOP algorithm (Eisenstein \& Hut 1998). Although it
remains popular, the FOF masses estimates are known to have
significant scatter due to a linking problem that can occur as small
strings of particles fall within the linking length.

The HOP algorithm works by using the local density for
each particle to trace (`hop') along a path of increasing density to
the nearest density maxima, at which point the particle is assigned to
the group defined by that local density maximum. As this process
assigns all particles to groups, a `regrouping' stage is needed in
which a merger criterion for groups above a threshold density
$\delta_{outer}$ is applied. This criterion merges all groups for
which the boundary density between them exceeds $\delta_{saddle}$, and
all groups thus identified must have one particle that exceeds
$\delta_{peak}$ to be accepted as a group (see Eisenstein \& Hut 1998
for explicit details).

Beginning from $z=4.89$, we saved particle positions every 50 million
years up to the final output at $z=3$. For the final 5 outputs we
found FOF groups using a linking parameter of $b=0.18$, and HOP groups
using the parameters: $N_{dens}=48$, $N_{hop}=20$, $N_{merge}=5$,
$\delta_{peak}=160$, $\delta_{saddle}=140$, and
$\delta_{outer}=80$. Visual inspection showed strong similarities
between the two halo populations, with a small amount of unavoidable
noise coming from groups around the 80 particle resolution limit (a
group found by FOF at this limit may not be found by HOP and vice
versa).  The group index of each particle was stored at each output to
enable tracing between outputs.

P03 investigated the clustering of mergers within simulations using the
FOF algorithm, finding no evidence for bias in a series of populations
defined using different selection criteria. To test their conclusions we
reconstructed their samples within our simulation. To give a rough
{\epsscale{0.93}
\plotone{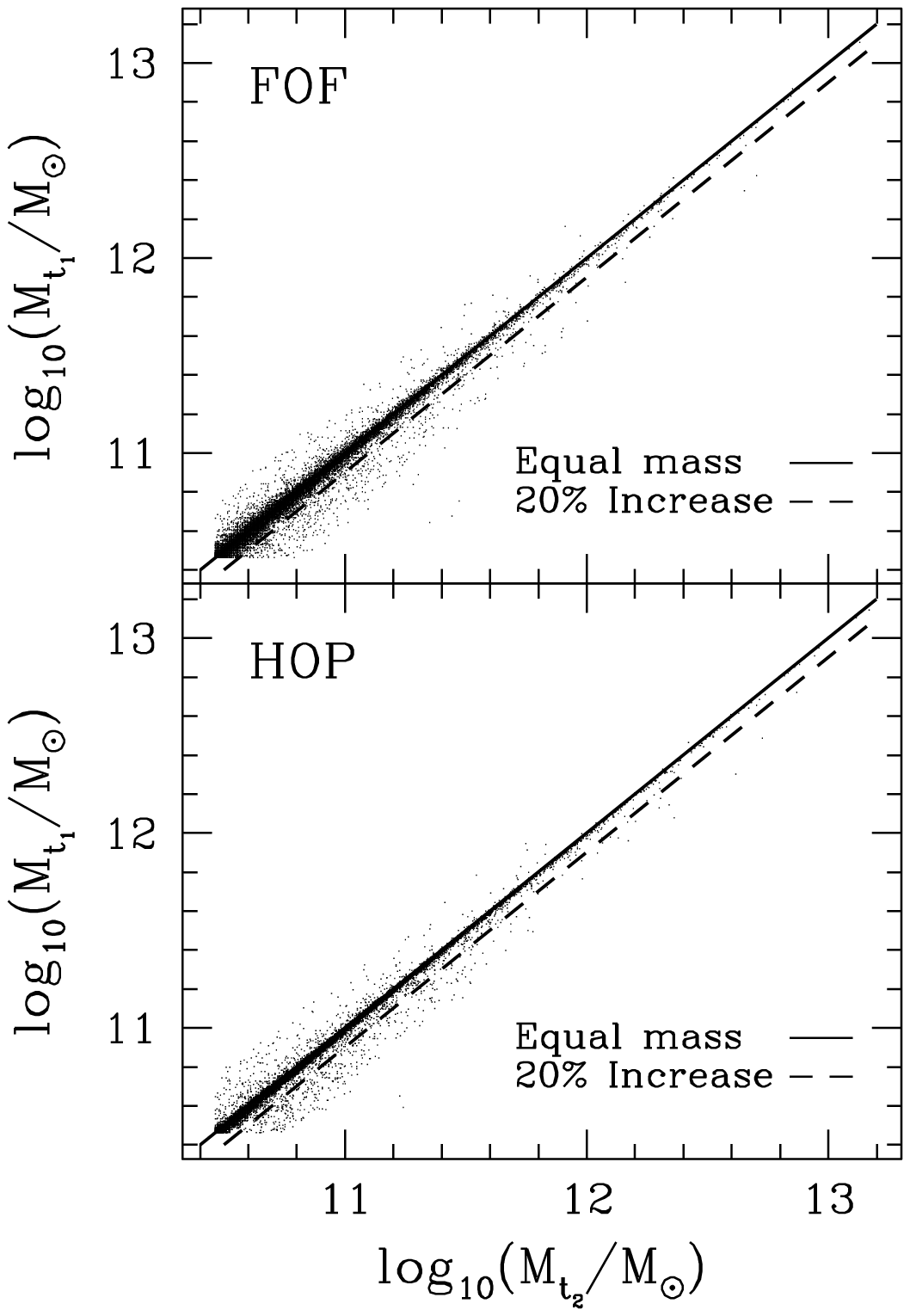}} {\small {\sc Fig.}~1.---Comparison of halo
growth. The FOF algorithm exhibits a significant amount of scatter in mass
estimates between outputs. Only 67\% of groups grow from one output to
another, compared to 
\medskip 82\% for HOP.\\} 
estimation of the accuracy
of the group finding methods in Fig.\ 1 we plot the mass of the most
massive progenitor at $t_1(z=3.059)$, versus the mass at $t_2(z=3)$,
such that $\Delta t = 5 \times 10^7$ yr.
Clearly HOP identifies groups that are more likely to be more massive at
later outputs, while FOF groups show considerable scatter about the mean.
The effect of this difference is significant. 

Using the P03 definition of `new' groups, namely, those for which 50\%
of constituent particles were not in a progenitor of equal or higher
mass at the previous time, we find a large difference between the
total number of groups identified: 9367 versus 12129 for FOF and HOP
respectively (from 16330 and 16831 total groups,
respectively). However, neither of these populations has stronger
clustering than the complete halo population as measured by their
correlation functions in mass bins of 0.5 dex.  Notably the P03 halo
subsets, corresponding to 3:1 and 17:3 mass ratio mergers,we
significantly different between the FOF and HOP populations.  This can
partially be attributed synchronization problems, and as the time
difference between outputs is increased the identified subsets become
more similar.  Further scatter is also introduced by hard selection
criteria, such as requiring progenitors fit a certain mass threshold.
For the 3:1 and 17:3 merger ratios we find only 171(226) groups
between $z=3.2$ and $z=3$, respectively. The correlation functions of
these halo populations, co-added over the final five outputs, show no
clear bias, confirming the conclusions of P03, however better
statistics are necessary for a definitive result.

\setcounter{figure}{1}
\begin{figure*}[t]
\vspace{120mm}
\includegraphics{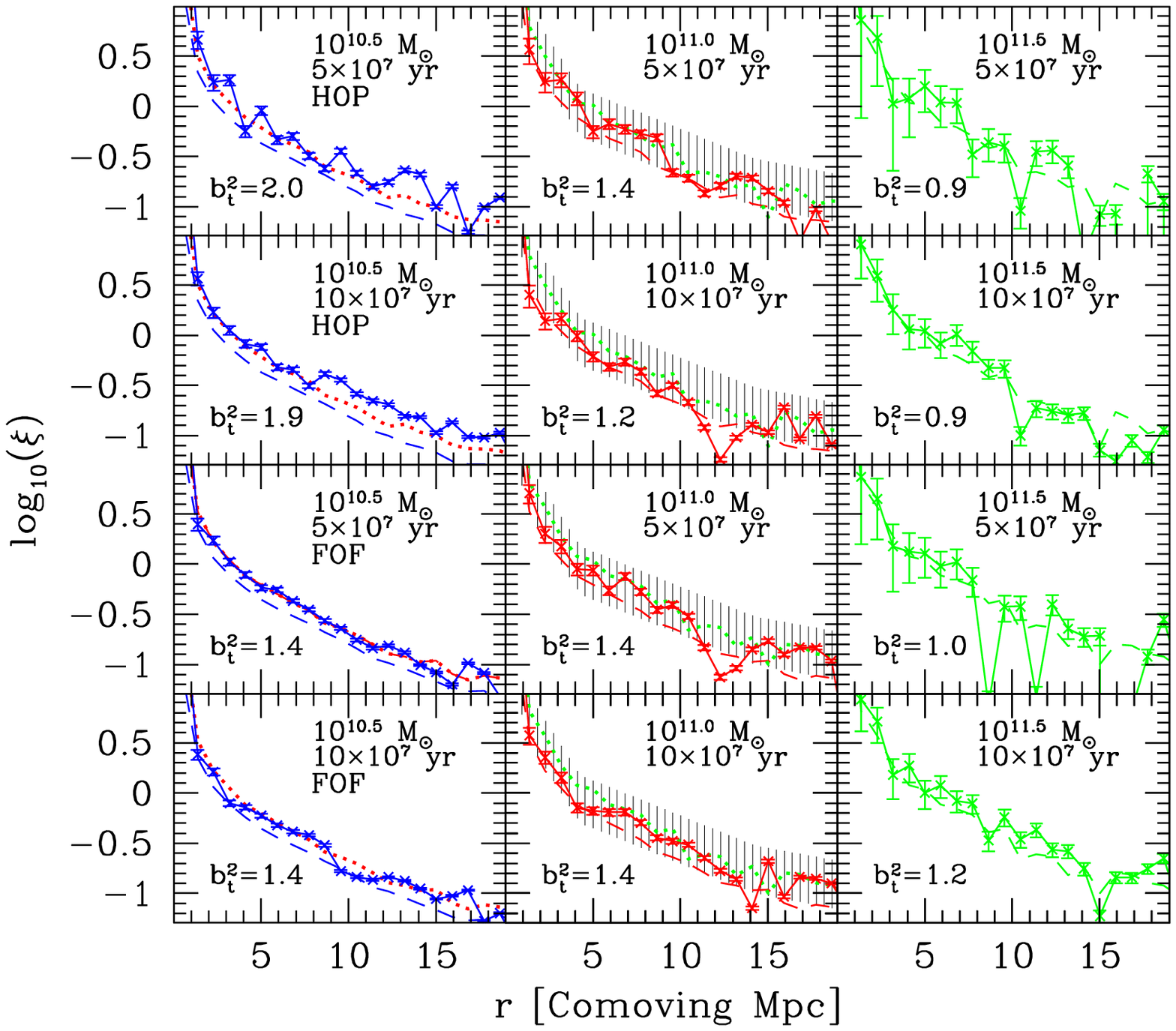}
\caption{Spatial Correlation Functions.
In each panel the dashed line shows the correlation function for all
the groups, which the points connected by the dashed lines show
$\xi(r)$ for groups that have accreted appreciable mass in the last
$\Delta t$ years.  Panels are labeled by their mass range and assumed
$\Delta t$ values, and in each panel, the dotted line shows the
correlation function of all the groups in the next highest mass bin.
The top two rows were generated from a set of groups selected by the
HOP algorithm, while the groups in the lower two rows were selected
using the FOF approach.  The shaded region in the central panels
represents the observed correlation function of ${\cal R}_{AB} \leq 25.5$
Lyman break galaxies as computed in Wechsler et al.\ 2001 by inversion
of the angular correlation function.  A 10\% accretion threshold is applied
in the $10^{11.5} \msun$ case to increase the number of measured groups.}
\label{fig:xi}
\end{figure*}

Our definition of accreting groups is similar to the P03 criteria,
except that we select the subset of halos that grew by 20\% from
output to output, which implicitly includes mass accretion via smooth
infall and results in 545(980) HOP(FOF) groups if $\Delta t = 5 \times
10^7.$ Note that the mass of each halo is that at the end of each time
interval, such that we tag all halos that {\em experienced}
appreciable infall.  The 20\% value is arbitrary, but we selected it
primarily because it appears to lie outside the central `noise' band
in the FOF data (see Fig\ 1). The number of halos corresponding to
this cut is also illustrated Fig\ 1, as points to the right of the
dashed lines.

\section{Temporal Bias}

In Figure \ref{fig:xi} we show the correlation function of the groups
selected by both the HOP and FOF algorithms and compare them with $\xi(r)$
of the accreting groups.  In the accreting case we co-added the
correlation functions calculated from the differences from the last
four $\Delta t = 5 \times 10^7$ year intervals and the last two $\Delta
t = 10 \times 10^7$ year intervals. Radial bins of 1/80 the simulation
size, corresponding to $0.92$ Mpc, were taken throughout.

For comparison, in each panel of Fig.\ \ref{fig:xi} we also show the
correlation function of all the groups in the next largest mass bin.
The $\nu \equiv 1.686 D(z)^{-1}\sigma(M)^{-1}$ values (where $D$ is
the linear growth factor) for each of these bins are 1.38, 1.58, and
1.83 which result in geometrical bias factors of $b \equiv
1+(\nu^2-1)/1.69$ of 1.53, 1.89, and 2.39 respectively.  The
amplitudes of the correlation functions obtained using all the HOP
groups are in good agreement with these bias values, and likewise, the
correlation functions of the full set of HOP and FOF groups agree with
each other to within statistical uncertainties.

The upper row of plots demonstrate a clear enhancement of the
clustering of accreting groups at both the $10^{10.5} M_\odot$ and
$10^{11.0} M_\odot$ mass scales, with their correlation functions
roughly matching those of objects three times greater in mass (no 
conclusion can be drawn from the high mass bin as the sample is too 
small). 
This ``temporal biasing'' arises from the fact that {\em both}
objects accreting substructure as well as those experiencing
considerable smooth infall tend to be found in the densest regions of
space, which are themselves highly clustered. This conclusion is supported 
by the fact that the average local overdensity of groups in the 
$10^{10.5}(10^{11}) 
M_\odot$ mass bin is 0.82(0.87) (measured in 4 Mpc (comoving) spheres, 
corresponding to a mass scale of $1.2\times10^{13}$ $M_\odot$), where as 
the 
same mass bin for the entire population exhibits an overdensity of 
0.60(0.73).

In the second row of Fig.\ \ref{fig:xi} we take a longer interval
of $\Delta t = 10 \times 10^7$ yr.  This has a slight dampening
effect on temporal bias, as groups in less dense regions are
able to sustain this level of infall.  Nevertheless a detectable
enhancement is still seen, particularly in the $ 10^{10.5} \msun$
bin, which contains the most common objects, with $\nu$ the
smallest.  In the $\Delta t = 20 \times 10^7$ yr case, however, only a
very weak enhancement of $\xi(r)$ was measured.

In the lower two rows of this figure, we repeat our analyses using the
FOF group finder.  Although this approach is more susceptible to
statistical noise, the same trends are apparent as in the HOP case.
If $\Delta t = 5 \times 10^7$ yr, this temporal bias is roughly equal
to the geometrical bias of the groups three times more massive, while
if $\Delta t = 10 \times 10^7$ yr, $\xi(r)$ is boosted to a slightly
lesser degree.

Besides the comparisons shown, we have also studied the effect of
varying a number of parameters: $\delta_{peak}$, $\delta_{saddle}$ and
$\delta_{outer}$ in the HOP method, the linking length of the FOF
groups, and the fractional value we used to define ``appreciable''
accretion.  Again, the same bias trends were visible, and our results
remained robust over a wide range of reasonable choices for these
quantities.

Finally, to quantify our results, we have computed the effective
temporal bias in each mass bin, $\Delta t$, and group finder.  We
define $b^2_t$ as the ratio of the correlation function of the
accreting groups to the overall correlation function, weighted by the
number of points in each bin in the overall function; $b^2_t \equiv
\sum^{20}_{i=0} \frac {\xi_{{\rm accreting},i} N_{{\rm
all},i}}{\xi_{{\rm all},i}N_{{\rm all},i}}$, where the sum is carried
out over all bins within $r \leq 20$ comoving Mpc.  These values are
labeled in each panel, and in the $\Delta t = 20 \times 10^7$ yr case,
$b^2_t =$ 1.1(1.0) in the $10^{10.5}(10^{11.0}) \msun$ HOP bins and 
1.1(1.3) in the respective FOF bins.

\section{Discussion}

From the tests presented above, it is clear that if small time
intervals are chosen, the clustering of accreting groups is robustly
enhanced with respect to the underlying populations. To relate this to
LBGs we plot the spatial correlation function of ${\cal R}_{AB} \leq
25.5$ LBGs, as derived by Wechsler et al.\ (2001), in the center
column of Fig.\ \ref{fig:xi}. Although there are significant
uncertainties involved in computing this quantity, since comparisons
are more naturally conducted in angular coordinates, the shaded
regions provide a guide to the range of $\xi(r)$ values consistent
with observations.  In these panels, we see that if $\Delta t = 5
\times 10^7$ yr is chosen, then temporal bias boosts the correlation
function of $10^{11} \msun$ halos into reasonable agreement with
observations.

This mass is marginally consistent with the upper mass bound inferred
from the rotation curves of a somewhat bright (${\cal R}_{AB} \lesssim
24$) spectroscopic subset of LBGs (Pettini et al.\ 2001).
Furthermore, only $\sim 4\%$ of all groups exhibit appreciable
accretion in each $\Delta t = 5 \times 10^7$ year time interval and
the density of $10^{11} \msun$ halos is $\sim 2 \times
10^{-2}$ Mpc$^3$, at $z = 3$ in our assumed cosmology.  Thus
associating such objects with $5 \times 10^7$ year starbursts results
in a density $\sim 5 \times 10^{-4}$ Mpc$^3$, comparable with that
observed.  While this time interval is small, taking the
mean gas and dark matter ratio, and an overall star formation
efficiency of $10\%$, results in a star formation rate of $0.2 \times
10^{11} \msun \Omega_b/\Omega_0/5 \times 10^7 \, {\rm yr}$ = $7\,
\msun$ yr$^{-1}$.  This is less than the observed values of $\sim 50
\msun$ yr$^{-1}$, perhaps implying that an even shorter time interval
is appropriate.

While quite suggestive, these comparisons are not meant as a complete
model, and may not prove to be the final explanation of the discrepant
mass estimates of LBGs.  Kinematic models have been explored, for
example, in which the observed velocity dispersions of LBGs are much
less than the circular velocities of the halos in which they are
contained (eg.\ Mo, Mao, \& White 1999).  What is clear however, is
that this bias can not be ignored and must be carefully considered
when interpreting the clustering of these objects.  While perhaps only
part of the story, temporal biasing represents an important factor
that must be taken into account when studying the properties of Lyman
break galaxies.

\acknowledgments 

ES would like to express his sincere thanks for the hospitality shown
to him by Jon Weisheit and the T-6 group at Los Alamos National
Laboratory, where this work was initiated.  We are grateful to Marc
Davis for fruitful suggestions and to Max Pettini for helpful comments.
ES was supported in part by an NSF MPS-DRF fellowship.  RJT\
acknowledges funding from the Canadian Computational Cosmology
Consortium and use of the CITA computing facilities.

\end{document}